\documentclass[
aps,
prl, 
twocolumn,
amsmath,
amssymb,
superscriptaddress
]{revtex4-2}

\usepackage{caption}
\newif\ifshowtodo
\usepackage{silence}
\usepackage[utf8]{inputenc}     % ソースをUTF-8として読む
\usepackage[T1]{fontenc}        % 欧文エンコーディング
\usepackage[whole]{bxcjkjatype} % ★日本語本文を組版できるようにする（CJK）

\usepackage{newtxtext}
\usepackage{newtxmath}

\usepackage{amsmath}
\usepackage{amssymb}
\usepackage{amsfonts}
\usepackage{bm}
\usepackage{mathtools}
\usepackage{pifont}

\usepackage{xcolor}
\usepackage{colortbl}

\usepackage{physics2}
\usephysicsmodule{ab}
\usephysicsmodule{ab.braket}
\usephysicsmodule{diagmat}
\usephysicsmodule{nabla.legacy}
\usephysicsmodule[showleft=2,showtop=2]{xmat}

\makeatletter
\newcommand{\vb}{\@ifstar\vb@star\vb@nostar}
\newcommand{\vb@star}[1]{\mathbf{#1}}
\newcommand{\vb@nostar}[1]{\boldsymbol{\mathit{#1}}}

\newcommand{\va}{\@ifstar\va@star\va@nostar}
\newcommand{\va@star}[1]{\vec{\mathrm{#1}}}
\newcommand{\va@nostar}[1]{\vec{#1}}

\newcommand{\vu}{\@ifstar\vu@star\vu@nostar}
\newcommand{\vu@star}[1]{\hat{\mathbf{#1}}}
\newcommand{\vu@nostar}[1]{\hat{\boldsymbol{#1}}}
\makeatother

\usepackage{fixdif}
\usepackage{derivative}

\usepackage{siunitx}
\usepackage{booktabs}
\usepackage{graphicx}
\usepackage{stackengine}

\usepackage{subfig} % \subfloat を使う

\usepackage{array}
\usepackage{here}
\usepackage{enumitem}

\newlist{todolist}{itemize}{2}
\setlist[todolist]{label=$\square$}

\usepackage{chngcntr}
\usepackage{tabularx}
\usepackage{tabularray}

\usepackage{etoolbox}
\ifshowtodo
  \usepackage[textsize=scriptsize]{todonotes}
\else
  \usepackage[disable]{todonotes}
\fi

\usepackage[version=3]{mhchem}
\usepackage{dirtree}
\usepackage{listings}
\usepackage{ascmac}
\usepackage{tcolorbox}
\usepackage{lineno}

\usepackage{indentfirst}

\usepackage[unicode,colorlinks=true, linkcolor=black, urlcolor=black, citecolor=black]{hyperref}
\usepackage{url}
\newcommand{\setPathToFigures}[1]{%
  \ifx\pathToFigures\undefined
    \newcommand{\pathToFiguresSubDir}{#1} % 画像ファイルのサブディレクトリを設定
    \newcommand{\pathToFigures}{figures/\pathToFiguresSubDir} % 画像パス全体を設定
  \else
    \renewcommand{\pathToFiguresSubDir}{#1} % サブディレクトリを再定義
    \renewcommand{\pathToFigures}{figures/\pathToFiguresSubDir} % 画像パス再定義
  \fi
}
\usepackage{siunitx}

\usepackage{tabularray}
\usepackage{mhchem}
\usepackage{graphicx}
\usepackage{subcaption}

\usepackage{lineno}
\begin{document}

\title{Probing In-Solid Proton Energy Distributions\\in Laser-Driven Fusion via Nuclear Activation Diagnostics}

\author{Hiroki Matsubara}
\email{u472377d@ecs.osaka-u.ac.jp}
\affiliation{Department of Physics, Graduate School of Science, The University of Osaka, 1-1 Machikaneyama-cho, Toyonaka, Osaka 560-0043, Japan}
\affiliation{Institute of Laser Engineering, The University of Osaka, 2-6 Yamada-Oka, Suita, Osaka 565-0871, Japan}
\author{Ryunosuke Takizawa}
\affiliation{Institute of Laser Engineering, The University of Osaka, 2-6 Yamada-Oka, Suita, Osaka 565-0871, Japan}
\author{Yuga Karaki}
\affiliation{Department of Physics, Graduate School of Science, The University of Osaka, 1-1 Machikaneyama-cho, Toyonaka, Osaka 560-0043, Japan}
\affiliation{Institute of Laser Engineering, The University of Osaka, 2-6 Yamada-Oka, Suita, Osaka 565-0871, Japan}
\author{Ryuya Yamada}
\affiliation{Department of Physics, Graduate School of Science, The University of Osaka, 1-1 Machikaneyama-cho, Toyonaka, Osaka 560-0043, Japan}
\affiliation{Institute of Laser Engineering, The University of Osaka, 2-6 Yamada-Oka, Suita, Osaka 565-0871, Japan}
\author{Tomoyuki Johzaki}
\affiliation{Graduate School of Advanced Science and Engineering, Hiroshima University, 1-3-2 Kagamiyama, Higashi-Hiroshima, Hiroshima, 739-8511, Japan}
\affiliation{Institute of Laser Engineering, The University of Osaka, 2-6 Yamada-Oka, Suita, Osaka 565-0871, Japan}
\author{Rinya Akematsu}
\affiliation{Department of Physics, Graduate School of Science, The University of Osaka, 1-1 Machikaneyama-cho, Toyonaka, Osaka 560-0043, Japan}
\affiliation{Institute of Laser Engineering, The University of Osaka, 2-6 Yamada-Oka, Suita, Osaka 565-0871, Japan}
\author{Ryo Omura}
\affiliation{Department of Physics, Graduate School of Science, The University of Osaka, 1-1 Machikaneyama-cho, Toyonaka, Osaka 560-0043, Japan}
\affiliation{Institute of Laser Engineering, The University of Osaka, 2-6 Yamada-Oka, Suita, Osaka 565-0871, Japan}
\author{Kai Kimura}
\affiliation{Department of Physics, Graduate School of Science, The University of Osaka, 1-1 Machikaneyama-cho, Toyonaka, Osaka 560-0043, Japan}
\affiliation{Institute of Laser Engineering, The University of Osaka, 2-6 Yamada-Oka, Suita, Osaka 565-0871, Japan}
\author{Fuka Nikaido}
\affiliation{Graduate School of Engineering, The University of Osaka, 2-1 Yamada-Oka, Suita, Osaka 565-0871, Japan}
\affiliation{Institute of Laser Engineering, The University of Osaka, 2-6 Yamada-Oka, Suita, Osaka 565-0871, Japan}
\author{Toshiharu Yasui}
\affiliation{Graduate School of Engineering, The University of Osaka, 2-1 Yamada-Oka, Suita, Osaka 565-0871, Japan}
\affiliation{Institute of Laser Engineering, The University of Osaka, 2-6 Yamada-Oka, Suita, Osaka 565-0871, Japan}
\author{Takumi Minami}
\affiliation{Graduate School of Engineering, The University of Osaka, 2-1 Yamada-Oka, Suita, Osaka 565-0871, Japan}
\affiliation{Institute of Laser Engineering, The University of Osaka, 2-6 Yamada-Oka, Suita, Osaka 565-0871, Japan}
\author{Law King Fai Farley}
\affiliation{Institute of Laser Engineering, The University of Osaka, 2-6 Yamada-Oka, Suita, Osaka 565-0871, Japan}
\author{Akifumi Yogo}
\affiliation{Institute of Laser Engineering, The University of Osaka, 2-6 Yamada-Oka, Suita, Osaka 565-0871, Japan}
\author{Yuki Abe}
\affiliation{Graduate School of Engineering, The University of Osaka, 2-1 Yamada-Oka, Suita, Osaka 565-0871, Japan}
\affiliation{Institute of Laser Engineering, The University of Osaka, 2-6 Yamada-Oka, Suita, Osaka 565-0871, Japan}
\author{Yasuhiro Kuramitsu}
\affiliation{Graduate School of Engineering, The University of Osaka, 2-1 Yamada-Oka, Suita, Osaka 565-0871, Japan}
\affiliation{Institute of Laser Engineering, The University of Osaka, 2-6 Yamada-Oka, Suita, Osaka 565-0871, Japan}
\author{Yuji Fukuda}
\affiliation{National Institutes for Quantum Science and Technology, Kansai Institute for Photon Science, 8-1-7 Umemidai, Kizugawa, Kyoto 619-0215, Japan}
\affiliation{Institute of Laser Engineering, The University of Osaka, 2-6 Yamada-Oka, Suita, Osaka 565-0871, Japan}
\author{Takehito Hayakawa}
\affiliation{National Institutes for Quantum Science and Technology, Kansai Institute for Photon Science, 8-1-7 Umemidai, Kizugawa, Kyoto 619-0215, Japan}
\affiliation{Institute of Laser Engineering, The University of Osaka, 2-6 Yamada-Oka, Suita, Osaka 565-0871, Japan}
\author{Masato Kanasaki}
\affiliation{Graduate School of Maritime Sciences, Kobe University, 5-1-1 Fukaeminami-machi, Higashinada-ku, Kobe 658-0022, Japan}
\affiliation{Graduate School of Engineering, The University of Osaka, 2-1 Yamada-Oka, Suita, Osaka 565-0871, Japan}
\author{Koichi Honda}
\affiliation{Institute of Laser Engineering, The University of Osaka, 2-6 Yamada-Oka, Suita, Osaka 565-0871, Japan}
\author{Kohei Yamanoi}
\affiliation{Institute of Laser Engineering, The University of Osaka, 2-6 Yamada-Oka, Suita, Osaka 565-0871, Japan}
\author{Keisuke Takahashi}
\affiliation{Institute of Laser Engineering, The University of Osaka, 2-6 Yamada-Oka, Suita, Osaka 565-0871, Japan}
\author{Koji Tsubakimoto}
\affiliation{Institute of Laser Engineering, The University of Osaka, 2-6 Yamada-Oka, Suita, Osaka 565-0871, Japan}
\author{Yu Yamamoto}
\affiliation{Blue Laser Fusion Inc., 6950 Hollister Ave, Goleta, CA 93117, United States}
\affiliation{Institute of Laser Engineering, The University of Osaka, 2-6 Yamada-Oka, Suita, Osaka 565-0871, Japan}
\author{Hideyuki Maruta}
\affiliation{Blue Laser Fusion Inc., 6950 Hollister Ave, Goleta, CA 93117, United States}
\affiliation{Institute of Laser Engineering, The University of Osaka, 2-6 Yamada-Oka, Suita, Osaka 565-0871, Japan}
\author{Atsushi Sunahara}
\affiliation{Blue Laser Fusion Inc., 6950 Hollister Ave, Goleta, CA 93117, United States}
\affiliation{Institute of Laser Engineering, The University of Osaka, 2-6 Yamada-Oka, Suita, Osaka 565-0871, Japan}
\author{Seita Iizuka}
\affiliation{Blue Laser Fusion Inc., 6950 Hollister Ave, Goleta, CA 93117, United States}
\affiliation{Institute of Laser Engineering, The University of Osaka, 2-6 Yamada-Oka, Suita, Osaka 565-0871, Japan}
\author{Shuji Nakamura}
\affiliation{Blue Laser Fusion Inc., 6950 Hollister Ave, Goleta, CA 93117, United States}
\author{Shinsuke Fujioka}
\email{fujioka.shinsuke.ile@osaka-u.ac.jp}
\affiliation{Institute of Laser Engineering, The University of Osaka, 2-6 Yamada-Oka, Suita, Osaka 565-0871, Japan}

\date{\today}

\begin{abstract}
The energy distribution of energetic protons inside a solid target is a key quantity governing nuclear reaction yields and energy deposition in high-intensity laser-driven fusion, including nonthermal proton--boron (p--B) schemes and proton fast ignition. 
Yet it has remained inaccessible to conventional particle diagnostics, which detect only ions escaping the target and are perturbed by intense plasma electromagnetic fields. 
Here we establish a quantitative diagnostic that uses nuclear activation reactions occurring within the target itself as an internal probe of the in-solid proton energy distribution. 
Applied to laser-driven p--B fusion experiments on the kJ-class laser, the method reconstructs an exponential-equivalent in-solid proton energy distribution from the absolute yields of $^{11}\mathrm{C}$ and $^{7}\mathrm{Be}$ produced via $\mathrm{^{11}B(p,n)^{11}C}$ and $\mathrm{^{10}B(p,\alpha)^{7}Be}$, and yields the absolute number of $\mathrm{^{11}B(p,2\alpha)^{4}He}$ reactions through a side-channel analysis with propagated cross-section uncertainties.
This work opens a quantitative window onto the in-solid proton dynamics that drive nuclear reactions in laser-driven fusion experiments.
\end{abstract}


\maketitle

A central problem common to laser-driven inertial fusion, fast ignition, laboratory astrophysics, and high-intensity laser-driven nuclear physics is the determination of the energy distribution of energetic ions \emph{inside} a dense medium. 
This in-solid distribution governs both the nuclear reaction yield and the spatial profile of energy deposition, and it is fundamentally distinct from that of externally detected ions escaping the target.

The recent achievement of ignition at the National Ignition Facility~\cite{PhysRevLett.132.065102} has intensified efforts toward practical fusion power~\cite{Hsu2023USFusion, Fujioka2024} and revived interest in nonthermal schemes~\cite{PhysRevE.72.026406, Labaune:2013aa, PhysRevX.4.031030, PhysRevE.103.053202, Margarone:2022aa} based on the proton--boron reaction, $\mathrm{^{11}B(p,2\alpha)^{4}He}$. 
These schemes, which exploit the low neutron yield and aneutronic character of p--B fusion~\cite{Labaune:2016aa}, depend critically on what fraction of the laser energy is deposited as fast protons inside the boron target---a question that current diagnostics cannot answer directly.

Conventional charged-particle diagnostics, such as the Thomson parabola (TP) ion energy analyzer and radiochromic film stacks~\cite{Carroll2010, Kantarelou2023, Nurnberg2009, Abe2021b}, sample only the small fraction of ions that escape the target. 
The escaping distribution is further distorted by sheath electric fields, self-generated megagauss magnetic fields~\cite{PhysRevLett.70.3059}, and self-absorption, none of which can be deconvolved without strong model assumptions. 
Diagnostics based on direct detection of the emitted $\alpha$ particles using solid-state track detectors~\cite{Cartwright1978, Cassou1978, Sonoda1983, Amemiya1988, Chen2011} are similarly compromised: track-size ambiguity between $\alpha$ particles and contaminant heavy ions~\cite{10.1063/1.4927684} and the field-induced redirection of emitted $\alpha$ particles introduce systematic uncertainties that are difficult to bound. 
As a result, the in-solid proton energy distribution---the very quantity that determines the fusion yield---has remained largely inaccessible.

\begin{figure}[t]
    \centering
    \includegraphics[width=1.0\linewidth]{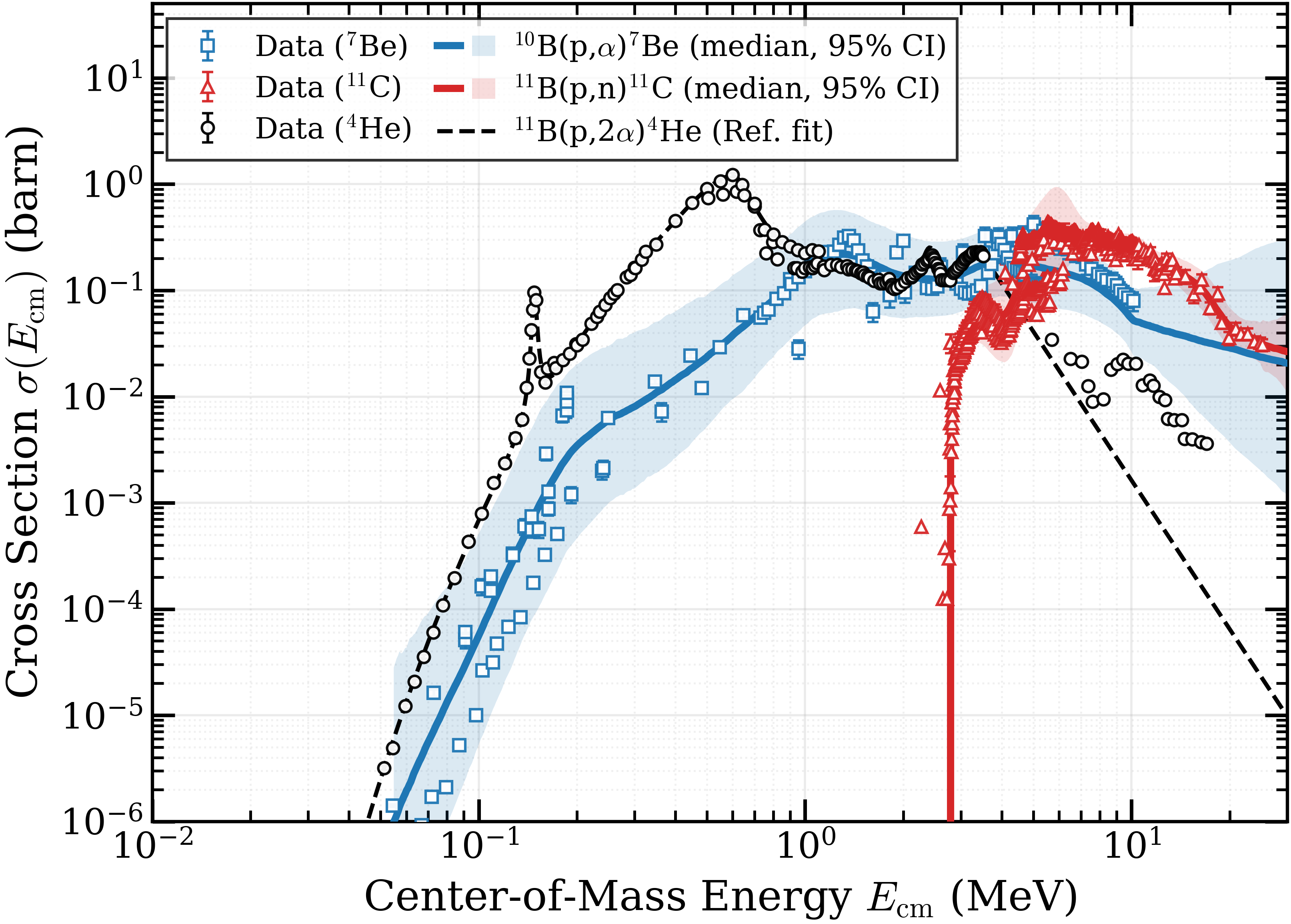}
    \caption{
Energy-dependent cross sections of the three p--B reaction channels relevant to this work (log--log scale).
Symbols show experimental data from EXFOR, supplemented for $\mathrm{^{10}B}(p,\alpha)\mathrm{^{7}Be}$ with measurements not yet registered in EXFOR~\cite{caciolli_2016, KAFKARKOU201348, PhysRevC.110.045806, PhysRevC.105.055802}; the complete dataset list is given in Appendix~B.
The solid curves and shaded bands show the posterior medians and 95\% credible intervals from the Bayesian analysis, while the dashed curve indicates the reference fit~\cite{Tentori_2023} adopted for $\mathrm{^{11}B(p,2\alpha)^{4}He}$.
    }
    \label{fig:proton-boron_cross_section}
\end{figure}

Here we develop and demonstrate a quantitative diagnostic that turns the target itself into a detector. 
Energetic protons traversing the boron target trigger the side channels $\mathrm{^{11}B(p,n)^{11}C}$ and $\mathrm{^{10}B(p,\alpha)^{7}Be}$, whose cross sections peak in different energy bands as shown in Fig.~\ref{fig:proton-boron_cross_section}. 
The absolute yields of the radioactive products $^{11}\mathrm{C}$ and $^{7}\mathrm{Be}$ therefore encode different moments of the proton energy distribution integrated along its path through the target. 
Measuring these yields by $\gamma$-ray spectroscopy of target debris collected after each shot and by solving the coupled proton-transport and nuclear-reaction problem, we reconstruct the in-solid proton energy distribution from internal nuclear signatures. 
The same reconstruction yields the absolute number of $\mathrm{^{11}B(p,2\alpha)^{4}He}$ reactions, providing a route to the fusion yield that is independent of direct $\alpha$-particle detection and free from the species-identification ambiguities intrinsic to track-detector methods.

%We apply this diagnostic to the kJ-class laser in two complementary geometries: a pitcher--catcher configuration where the activation result can be cross-checked against a conventional TP, and an in-target configuration where it cannot. 
%Agreement in the former validates the method, while in the latter the diagnostic remains operational in a regime inaccessible to external particle analyzers. 
%The conceptual framework---using selected nuclear side channels as built-in energy analyzers---is broadly applicable to laser-driven nuclear physics, including energy-deposition diagnostics for proton fast ignition~\cite{Roth2001FastIgnition, Fernandez2014FastIgnition}, laboratory nuclear astrophysics, and high-intensity ion-beam material science.

We apply this diagnostic to the kJ-class laser in two complementary geometries: a pitcher--catcher configuration where the activation result can be cross-checked against a conventional TP, and an in-target configuration where it cannot. 
Agreement in the former validates the method, while in the latter the diagnostic remains operational in a regime inaccessible to external particle analyzers. 
The conceptual framework---using selected nuclear side channels as built-in 
energy analyzers---is broadly applicable to laser-driven nuclear physics, 
including energy-deposition diagnostics for proton fast ignition~\cite{Roth2001FastIgnition, Fernandez2014FastIgnition}, 
laboratory nuclear astrophysics, and high-intensity ion-beam material science.
An extended discussion of the physical motivation for laser-driven p--B fusion 
and the systematic limitations of existing diagnostics is provided in 
Appendix~A.

The cross sections of the three p--B reaction channels relevant to this work are shown in Fig.~\ref{fig:proton-boron_cross_section}.
Across the analysis range ($E_{\mathrm{cm}}=0.01$--$30$~MeV), uncertainties in these cross sections propagate directly to the proton-distribution parameters $(T_{0}, A_{0})$ inferred from the activation analysis.
We therefore evaluate each cross section as the posterior of a Bayesian model for $\log\sigma(E)$ (Appendix~B) so that these uncertainties enter the final result through error propagation.
The data come primarily from EXFOR, supplemented for $\mathrm{^{10}B}(p,\alpha)\mathrm{^{7}Be}$ by recent measurements not yet registered in the database~\cite{caciolli_2016, KAFKARKOU201348, PhysRevC.110.045806, PhysRevC.105.055802}; these recent measurements significantly tighten the constraint in the $E_{\mathrm{cm}}=0.2$--$1$~MeV region, where EXFOR data alone are sparse and the posterior would otherwise be markedly broader.
For $\mathrm{^{11}B}(p,n)\mathrm{^{11}C}$, the substantial dispersion among reported values in the 4--10~MeV region is similarly absorbed into the posterior.
The $\mathrm{^{11}B(p,2\alpha)^{4}He}$ channel does not enter the measured radioisotope yields and is used only to evaluate the final fusion reaction number $Y_{\alpha}$; for this channel we adopt the reference fit from a published evaluation~\cite{Tentori_2023}.
In Fig.~\ref{fig:proton-boron_cross_section}, the solid curves and shaded bands show the posterior medians and 95\% credible intervals; the dashed curve is the reference fit.

The two absolute yields of $^{11}\mathrm{C}$ and $^{7}\mathrm{Be}$ obtained
from the activation measurement constitute two independent observables of the
in-solid proton energy distribution, restricting the reconstruction to two degrees of freedom.
We model the incident proton energy distribution at the boron-containing 
target with the two-parameter Boltzmann form
\begin{equation}
f(E_{0};A_0, T_0) = A_{0}\exp\!\left(-E_{0}/T_{0}\right), \qquad E_{\mathrm{min}} \le E_{0} \le E_{\mathrm{cut\text{-}off}},
\label{eq:Boltzmann}
\end{equation}
where $E_{0}$, $T_{0}$, and $A_{0}$ are respectively energy, slope temperature, and absolute normalization of the incident protons on the boron target surface, and $E_{\mathrm{cut\text{-}off}}$ the maximum energy characteristic of the acceleration mechanism.
This single-temperature Boltzmann form is well established for TNSA, both experimentally and theoretically~\cite{Zimmer2021,Macchi:2013aa}.
We have also tested two other two-parameter forms---a power-law form,
frequently used in the stochastic-acceleration regime, and the analytic form
derived for isothermal expansion of a plasma into vacuum~\cite{PhysRevLett.90.185002}; the Boltzmann form yielded the smallest residual against the activation data in all shots, and is adopted throughout this work.

Inside the target, the proton energy at depth $x$ evolves as
\begin{equation}
E(x) = E_{0} - \int_{0}^{x} S\!\left(E(x')\right)\,\mathrm{d}x',
\label{eq:slowing_down}
\end{equation}
where $S(E)$ is the SRIM-2013 stopping power~\cite{ZIEGLER20101818}.
The cut-off energy $E_{\mathrm{cut\text{-}off}}$ is determined for each shot by inverting Eq.~(\ref{eq:slowing_down}) on the maximum proton energy observed by the TP placed behind the target in both configurations.

The yield of side channel $j$ is then
\begin{equation}
Y_{j} = \int_{E_{\mathrm{min}}}^{E_{\mathrm{cut\text{-}off}}} \!\!\mathrm{d}E_{0}\, f(E_{0})\, n_{i}\!\int_{E(d)}^{E_{0}}\frac{\sigma_{j}(E)}{S(E)}\,\mathrm{d}E,
\label{equation:Yield}
\end{equation}
where $n_{i}$ is the number density of the relevant target nucleus, $d$ the target thickness, and $E_{\mathrm{min}}=0.01$~MeV the lower integration bound.
$T_{0}$ is then determined uniquely from the yield ratio $Y_{^{7}\mathrm{Be}}/Y_{^{11}\mathrm{C}}$, which is independent of $A_{0}$, and $A_{0}$ is fixed by a weighted least-squares fit to the two absolute yields.
Two sources of uncertainty are propagated into the inferred $(T_{0}, A_{0})$ by Monte Carlo sampling: the Bayesian posteriors of the cross sections, and the systematic uncertainty in the SRIM stopping power, taken as the mean bias $+0.6\%$ and standard deviation $7.1\%$ reported by the IAEA stopping power database for protons in solid compounds~\cite{IAEA_stopping_power_database}.

Experiments were performed with the kJ-class laser facility (central wavelength 1053~nm, typical pulse energy 1~kJ, duration 1.3~ps, peak focused intensity $5\times10^{18}$~W/cm$^{2}$); per-shot parameters are summarized in Appendix~C. 
Two boron-containing targets, decaborane ($\mathrm{B_{10}H_{14}}$) and borophane (BH)~\cite{Nishino2017JACS}, both 1~mm thick and of natural isotopic composition, were used. 
Decaborane targets, which sublimate in vacuum, were sealed with a 20~$\mu$m CH film on the laser-incidence side and epoxy elsewhere.

\begin{figure}[t]
\centering
\includegraphics[width=1.0\linewidth]{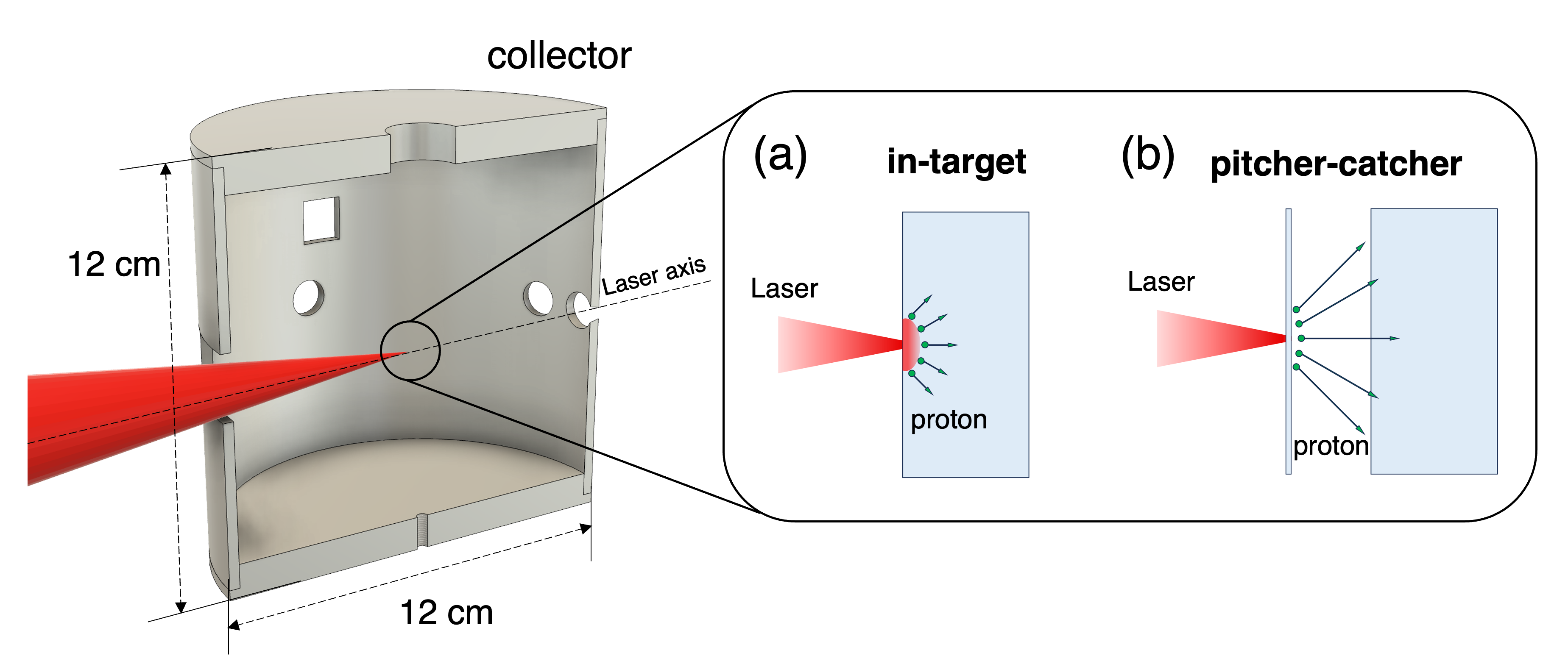}
\caption{
Schematic of the two experimental configurations. (a)~in-target: the laser is focused directly on a boron-containing target. (b)~pitcher--catcher: TNSA protons from a polypropylene pitcher foil traverse a vacuum gap and impinge on a separate boron catcher. In both cases the target is enclosed by a cylindrical aluminum debris collector lined with high-purity foil that captures the radioactive ejecta.
}
\label{fig:experiment_setup}
\end{figure}

Two geometries were investigated (Fig.~\ref{fig:experiment_setup}). 
In the \emph{in-target} configuration, the laser was focused directly onto the boron target, and protons originating from surface contaminants were accelerated into the target by ponderomotive and hole-boring forces~\cite{Macchi:2013aa,Robinson2009HoleBoringPPCF}. 
In the \emph{pitcher--catcher} configuration, a 10~$\mu$m polypropylene pitcher foil generated TNSA protons~\cite{Macchi:2013aa} that crossed a vacuum gap before striking the boron catcher. 
Three shots were taken in each geometry, with combinations of decaborane and borophane targets.

The radioactive products $\mathrm{^{11}C}$ and $\mathrm{^{7}Be}$ are released from the target as part of the laser-driven debris ejected over a wide solid angle.
They were captured on a removable high-purity aluminum foil lining a cylindrical collector (12~cm diameter, 12~cm height; 4\% aperture for laser injection and diagnostics, giving an isotropic-emission recovery efficiency of 96\%) surrounding the target. 
After each shot, the foil was transferred to a low-background HPGe spectroscopy station (see Appendix~C).

The 511~keV peak contains, in addition to $^{11}\mathrm{C}$, contributions from two $\beta^{+}$-emitting impurity nuclides that share the same annihilation $\gamma$ line: $^{34\mathrm{m}}\mathrm{Cl}$ ($T_{1/2}=32.0$~min)~\cite{Vandecasteele:1980aa,Lagunas-Solar:1992aa}, originating from chlorine and sulfur contamination, and $^{13}\mathrm{N}$ ($T_{1/2}=9.97$~min), produced by proton reactions with surface C/N/O contaminants.
We therefore performed a three-component exponential fit to separate the contribution of each nuclide (see Appendix~C).
Because the abundance of $^{34\mathrm{m}}\mathrm{Cl}$ can be estimated independently from its other characteristic $\gamma$-ray peaks, this estimate was used as the initial guess for the fit.
The resulting $^{11}\mathrm{C}$ fraction $f_{^{11}\mathrm{C}}$ depends on the configuration: 0.9--1.0 in the pitcher--catcher case and 0.3--0.4 in the in-target case.
The 478~keV peak was well described by a single exponential whose fitted half-life ($53.1\pm2.7$~days) is consistent with the $^{7}\mathrm{Be}$ literature value (53.2~day~\cite{ENSDF}). 
The debris distribution on the collector was imaged by autoradiography; the corresponding position-dependent HPGe detection efficiency, was computed by Geant4 Monte Carlo~\cite{Agostinelli:2003aa} and calibrated with $^{152}\mathrm{Eu}$ and $^{133}\mathrm{Ba}$ standard sources~\cite{Sarasti:2022}.
Full procedural details and decay-curve examples are given in Appendix~C.

Table~\ref{tab:summary_results} summarizes the absolute yields, the reconstructed parameters $T_{0}$ and $A_{0}$, the inferred number of $\mathrm{^{11}B(p,2\alpha)^{4}He}$ reactions, and the deduced laser-to-proton conversion efficiency $\xi$ for all six shots. 

\begin{table*}[t]
\centering
\caption{
 Summary of measured radioisotope yields, reconstructed proton-beam parameters, and inferred $\mathrm{^{11}B(p,2\alpha)^{4}He}$ reaction numbers ($Y_{\alpha}$ refers to the total $\alpha$-particle yield, three per fusion event). 
 $\xi$ is the laser-to-proton conversion efficiency. 
$N_{0}(^{11}\mathrm{C})$ and $N_{0}(^{7}\mathrm{Be})$ are scaled by $10^{7}$, and $N_{0}(\alpha)$ is scaled by $10^{9}$.
}
\label{tab:summary_results}
\scalebox{1.0}{
\begin{tblr}{
    colspec = {Q[l,m] Q[c,m] Q[c,m] Q[c,m] Q[c,m] Q[c,m] Q[c,m] Q[c,m] Q[c,m]},
    colsep  = 5pt,
    rows    = {font=\footnotesize, valign=m},
    row{1}  = {font=\footnotesize\bfseries},
    hline{1,Z} = {0.8pt},
    hline{2}   = {0.5pt},
}
Shot ID
& {$E_{\mathrm{cut\text{-}off}}$ \\ $[\mathrm{MeV}]$}
& $f_{^{11}\mathrm{C}}$
& {$N_0(^{11}\mathrm{C})$ \\ $[\times10^{7}\,\mathrm{atoms}]$}
& {$N_0(^{7}\mathrm{Be})$ \\ $[\times10^{7}\,\mathrm{atoms}]$}
& {$T_0$ \\ $[\mathrm{MeV}]$}
& {$A_0$ \\ $[\times10^{13}\,\mathrm{protons/MeV}]$}
& {$N_0(\alpha)$ \\ $[\times10^{9}\,\mathrm{atoms}]$}
& {$\xi$ \\ $[\%]$}
\\ \hline
\SetCell[c=9]{halign=l, font=\footnotesize\bfseries} in-target configurations &&&&&&&& \\
L5944 & $11$ & $0.31 \pm 0.11$ & $0.41 \pm 0.14$ & $3.5 \pm 0.5$  & $0.70_{-0.03}^{+0.03}$ & $3.9_{-2.1}^{+2.7}$  & $1.4_{-0.7}^{+1.6}$  & $0.40_{-0.22}^{+0.25}$ \\
L5947 & $15$ & $0.30 \pm 0.21$ & $0.13 \pm 0.09$ & $2.4 \pm 0.3$  & $0.59_{-0.03}^{+0.02}$ & $4.7_{-2.5}^{+3.5}$  & $1.2_{-0.6}^{+1.4}$  & $0.48_{-0.27}^{+0.32}$ \\
L5949 & $11$ & $0.43 \pm 0.14$ & $0.50 \pm 0.17$ & $2.5 \pm 0.2$  & $0.81_{-0.03}^{+0.03}$ & $1.7_{-0.9}^{+1.1}$  & $0.85_{-0.42}^{+0.92}$ & $0.32_{-0.17}^{+0.22}$ \\
\hline
\SetCell[c=9]{halign=l, font=\footnotesize\bfseries} pitcher--catcher configurations &&&&&&&& \\
L5941 & $17$ & $0.97 \pm 0.02$ & $43 \pm 1$ & $64 \pm 1$ & $1.16_{-0.06}^{+0.06}$ & $13_{-6}^{+9}$ & $16_{-8}^{+16}$ & $4.0_{-2.0}^{+2.6}$ \\
L5945 & $21$ & $1.00 \pm 0.01$ & $63 \pm 1$ & $26 \pm 1$ & $2.2_{-0.3}^{+0.3}$    & $0.92_{-0.49}^{+0.65}$ & $3.8_{-1.8}^{+3.8}$ & $0.93_{-0.46}^{+0.69}$ \\
L5948 & $23$ & $0.89 \pm 0.01$ & $102 \pm 2$ & $53 \pm 1$ & $1.8_{-0.2}^{+0.2}$  & $2.7_{-1.4}^{+1.9}$ & $8.9_{-4.2}^{+8.7}$ & $2.8_{-1.5}^{+2.0}$ \\
\end{tblr}
}
\end{table*}

\begin{figure}[t]
    \centering
    \includegraphics[width=1\linewidth]{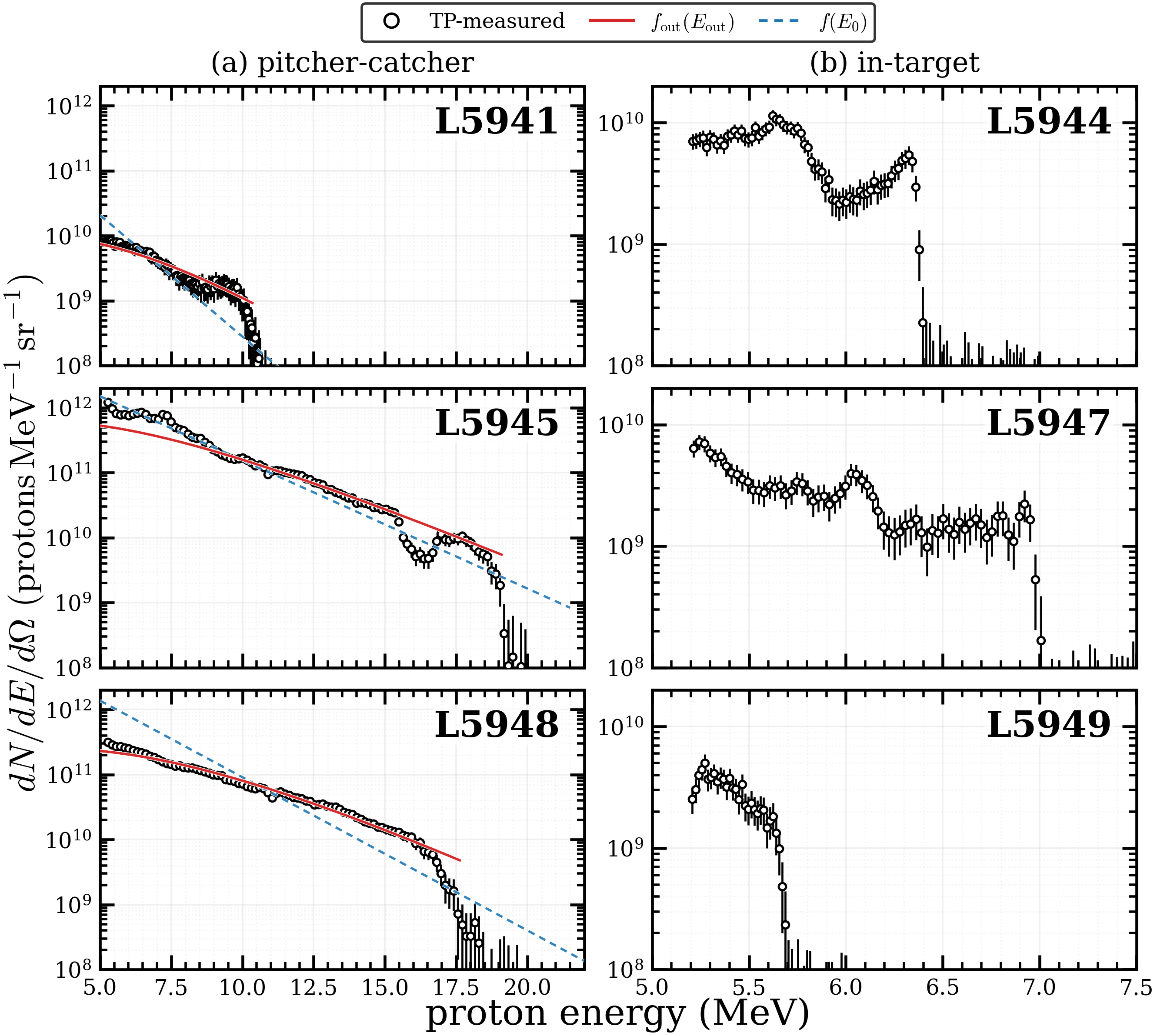}
   \caption{
Proton energy distributions in the (a)~pitcher--catcher and (b)~in-target configurations.
In both (a) and (b), open circles show the TP-measured distributions.
In~(a), the blue dashed and red solid curves show, respectively, the activation-derived incident distributions $f(E_0)$ and the same distributions after forward-propagation through the 1~mm boron target, $f_{\mathrm{out}}(E_{\mathrm{out}})$.
To enable a spectral-shape comparison, $f(E_0)$ and $f_{\mathrm{out}}(E_{\mathrm{out}})$ curves are adjusted to TP-measured distributions using a log-space least-squares fit.
The TP-measured distributions and $f_{\mathrm{out}}(E_{\mathrm{out}})$ curves's shape agree above $\sim 5.2$~MeV, validating the activation method.
    }
    \label{fig:compare_Tp_activation_TP}
\end{figure}

\emph{Validation in the pitcher--catcher geometry.}---We cross-check the activation result against an independent TP placed on the rear side of the catcher.
The two diagnostics probe different but related quantities: activation responds to the proton distribution \emph{incident} on the catcher, $f(E_{0})$, while the TP measures the distribution \emph{transmitted} through it, $f_{\mathrm{out}}(E_{\mathrm{out}})$.
The raw TP-measured distributions for both configurations are shown in Fig.~\ref{fig:compare_Tp_activation_TP}; the pitcher--catcher distributions are well described by a Boltzmann form, while the in-target distributions are not (see below).

To bridge the two, we forward-propagate $f(E_{0})$ through the 1-mm-thick boron target using the same SRIM stopping power $S(E)$ as in the main analysis. 
The incident and transmitted energies are related by
\begin{equation}
    d = \int_{E_{\mathrm{out}}}^{E_{0}} \frac{\mathrm{d}E}{S(E)},
\end{equation}
The transmitted distribution is then obtained from particle conservation,
\begin{equation}
    f_{\mathrm{out}}(E_{\mathrm{out}}) 
    = f\!\left(E_{0}(E_{\mathrm{out}})\right)
      \left|\frac{\mathrm{d}E_{0}}{\mathrm{d}E_{\mathrm{out}}}\right|.
\end{equation}
Here, we neglect proton depletion by nuclear reactions during transport.
This approximation is justified a posteriori because the reacted-proton fraction is at most $\sim 10^{-4}$ for all shots (Table~\ref{tab:summary_results}).

Figure~\ref{fig:compare_Tp_activation_TP} (a) compares the propagated distribution $f_{\mathrm{out}}(E_{\mathrm{out}})$ (red solid curves) and $f(E_{0})$ (blue dashed curves) with the TP-measured distribution (open circles) for the three pitcher--catcher shots.
To enable a spectral-shape comparison, $f(E_0)$ and $f_{\mathrm{out}}(E_{\mathrm{out}})$ curves are adjusted to TP-measured distributions using a log-space least-squares fit.
$f_{\mathrm{out}}(E_{\mathrm{out}})$ and the TP-measured distributions are in good agreement across the energy range covered by the TP above 5.2~MeV.

This consistency confirms that the activation method correctly infers the effective in-solid proton energy distribution in the regime where an independent benchmark is available. 
Moreover, the agreement implies that the cold-matter SRIM stopping power remains a reasonable approximation for this transport calculation, despite the heating that the energetic proton bombardment may induce in the target.

\emph{Application to the in-target geometry.}---The same TP cross-check is not meaningful in the in-target case.
The activation analysis indicates that the in-target proton distribution reaches the boron with a slope temperature a factor of two-to-four lower than in the pitcher--catcher geometry because, in this geometry, the protons receive no additional sheath-field boost at the rear surface, in contrast to the
pitcher-catcher case.
Within the limited high-energy window accessible to the TP [Fig.~\ref{fig:compare_Tp_activation_TP}(b)], only the tail of this distribution survives, and that tail is heavily distorted by the sheath; a single-Boltzmann fit is therefore not justified.
The activation diagnostic, by contrast, integrates over the entire proton population that crosses the boron and is unaffected by the sheath geometry, so its result is not biased by the same systematics that disqualify the TP.
The two methods are thus complementary: in regimes where an external benchmark exists they agree, and in regimes where it does not, the activation diagnostic provides a quantitative internal probe onto the in-solid distribution---the very quantity needed to interpret the fusion yield in geometries relevant to fast ignition and in-volume p--B fusion.

One concern that deserves explicit examination is the uncertainty in the stopping power model inside the directly laser-irradiated boron target, where the bulk material can be heated and partially ionized by the laser--plasma interaction.
To bound this effect, we performed Monte Carlo sampling of $S(E)$ with an artificially inflated standard deviation of $20\%$, well above the $7.1\%$ cold-matter value~\cite{IAEA_stopping_power_database} adopted in the main analysis, and propagated it through the activation reconstruction.
The induced deviations in the inferred parameters are $2\%$ for $T_{0}$ and $10\%$ for $A_{0}$.
We conclude that $T_{0}$ is essentially insensitive to the stopping-power modeling, while $A_{0}$ is only modestly affected.

\begin{figure}[ht]
    \centering
    \includegraphics[width=1.0\linewidth]{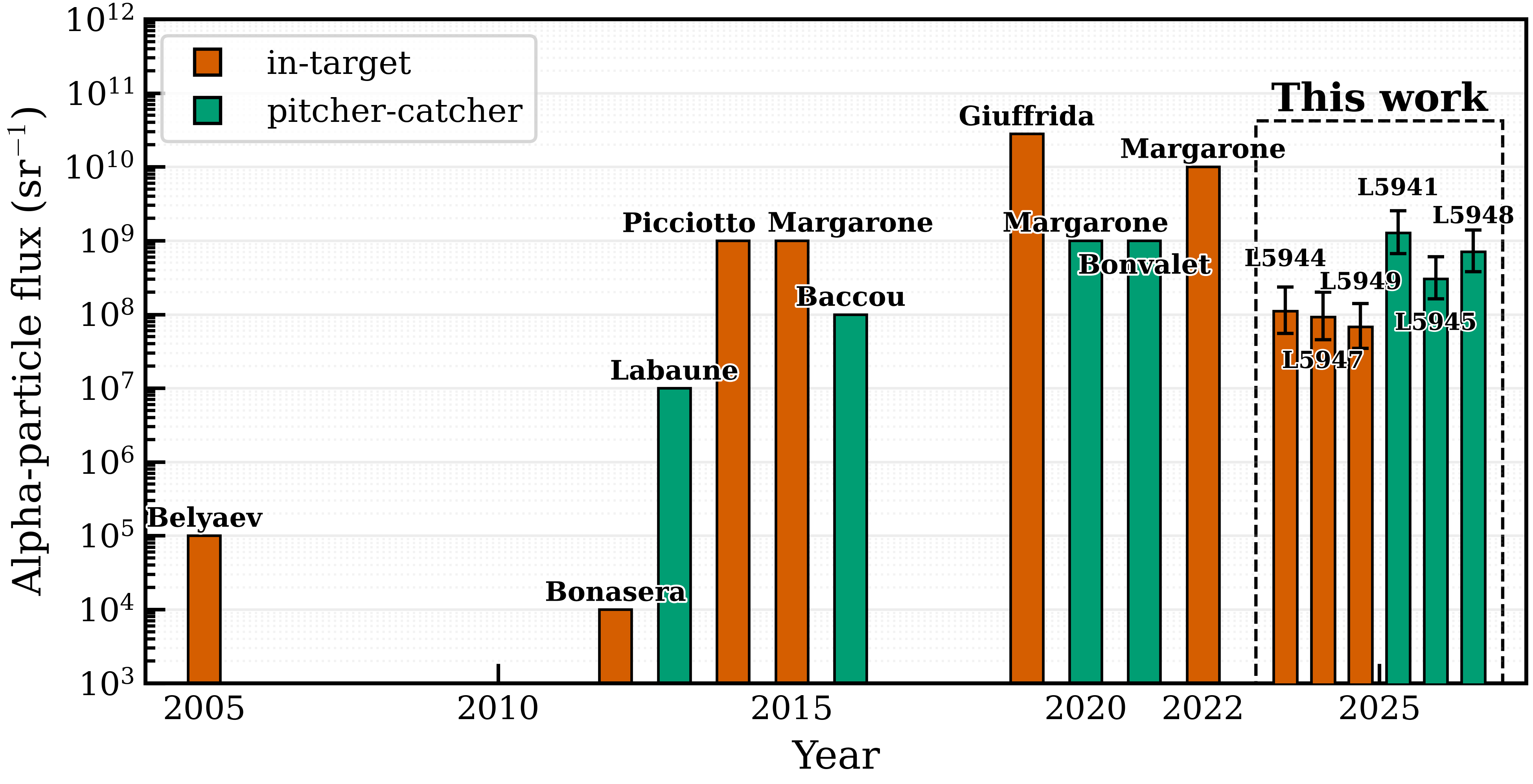}
    \caption{
Comparison of $\alpha$-particle yields per unit solid angle from the p--B reaction obtained in this study and summarized in Ref.~\cite{Margarone:2022aa}. 
The present values are derived from activation-based reconstruction of the in-solid proton distribution under the assumption of isotropic $\alpha$ emission.}
    \label{fig:alpha_yield}
\end{figure}

\emph{Reaction yields and comparison with the literature.}---The total numbers of $\mathrm{^{11}B(p,2\alpha)^{4}He}$ reactions inferred for the six shots, expressed as $\alpha$-particle yields per unit solid angle in Fig.~\ref{fig:alpha_yield}, are comparable to recent CR-39 values~\cite{Margarone:2022aa}.
The comparison is significant because, under the present experimental conditions, a direct CR-39 measurement performed in parallel could not unambiguously discriminate $\alpha$-particle tracks from those produced by co-accelerated heavier ions; the activation diagnostic recovers the absolute reaction number despite this limitation, by virtue of its species-specific nuclear signatures.

In summary, we have established a quantitative diagnostic of the in-solid proton energy distribution in laser-driven fusion experiments, using nuclear activation reactions.
Validation in the pitcher--catcher geometry confirms its accuracy against an independent TP measurement, while application in the in-target geometry---where conventional charged-particle diagnostics fail---provides access to the exponential-equivalent in-solid distribution and the absolute number of $\mathrm{^{11}B(p,2\alpha)^{4}He}$ reactions.
Beyond p--B fusion, the same framework, in which selectable nuclear side reactions serve as built-in energy analyzers, is readily portable to other laser-driven nuclear physics experiments, including fast-ignition energy-deposition diagnostics, laboratory studies of explosive nucleosynthesis, and high-intensity ion-beam material science.
Realizing its full quantitative potential will require renewed precision measurements of the relevant low-energy nuclear cross sections; recent measurements of the $\mathrm{^{10}B(p,\alpha)^{7}Be}$ cross section have already tightened the constraint in the $E_{\mathrm{cm}} = 0.2 - 1$ MeV region, where EXFOR data alone are sparse and the posterior would otherwise be markedly broader.

\begin{acknowledgments}
The authors thank the technical support staff for their assistance with operation, target fabrication, plasma diagnostics, and computer simulations.
This work was supported by the Joint Usage/Research Center Program of the Institute of Laser Engineering (ILE) at The University of Osaka (OU); the Collaborative Research Program between the National Institute for Fusion Science and ILE; Grants-in-Aid for Scientific Research (Nos.~25K17369, 23K25847, 23K03360, 22H00118, 22H01205, 22H01206, 21H04454, 20H00140, and 20H01886); the ``Power Laser DX Platform'' as shared research equipment under the Ministry of Education, Culture, Sports, Science and Technology (MEXT) Project for Promoting Public Utilization of Advanced Research Infrastructure (Program for Advanced Research Equipment Platforms, Grant No.~JPMXS0450300024); the ``OU Core Facilities'' as the MEXT Program for Supporting Construction of Core Facilities (Grant No.~JPMXS0441200024); and the Japan Society for the Promotion of Science Core-to-Core Program (Grant No.~JPJSCCA20230003).
The OU Honors Program for Graduate Schools in Science, Engineering and Informatics partially supported H.~M., Y.~K., and Y.~Y.
This work was further supported by grants conducted under the framework of the Blue Laser Fusion Energy Research Alliance Laboratory at the University of Osaka.
During preparation of this manuscript, the authors used LLM to improve language clarity and readability; all content was subsequently reviewed and edited by the corresponding authors, who take full responsibility for the final version.
\end{acknowledgments}

% =======================================================================
% Appendices
% (Former Supplemental Material: Sec.~SM-1, SM-2, SM-3 are now Appendices A,
% B, and C, respectively. Cross-references to "Sec.~SM-N of the Supplemental
% Material" in the main text have been redirected to these appendices,
% which are referenced as "Appendix~A", "Appendix~B", and "Appendix~C".)
% =======================================================================
\appendix

% Force appendix section numbering to A, B, C... so the printed headings
% read "Appendix A:", "Appendix B:", "Appendix C:".
% Note: in REVTeX/PRL, \section under \appendix does not produce a printed
% number, so \ref{...} returns an empty string. Therefore the main-text
% citations are written explicitly as "Appendix~A", "Appendix~B", and
% "Appendix~C" rather than via \ref. The \label{...} commands are kept
% for completeness and for any downstream tooling that resolves them.
\renewcommand{\thesection}{\Alph{section}}
\renewcommand{\thesubsection}{\thesection.\arabic{subsection}}

% -----------------------------------------------------------------------
% Appendix A (former Sec. SM-1)
% -----------------------------------------------------------------------
\section{Appendix A: Extended discussion of p--B fusion and existing diagnostics}
\label{sec:SM_background}

The proton--boron reaction $\mathrm{^{11}B(p,2\alpha)^{4}He}$ has long attracted attention as an aneutronic fusion candidate. The associated neutron yield from secondary channels is below the percent level~\cite{Labaune:2016aa}, dramatically reducing radiation damage and long-lived activation of reactor structures. Hydrogen and boron are nonradioactive, can be handled in solid form at room temperature, and require no cryogenic infrastructure.

The corresponding physics, however, is unforgiving. The cross section shows compound-nucleus resonances at $E_{\mathrm{cm}}\approx 148$~keV and 614~keV~\cite{Ajzenberg-Selove:1990aa}, the latter reaching $\sim 1$~barn---roughly six times the energy of the D--T peak and about one-sixth of its peak magnitude. The higher mean atomic number of a p--B plasma also amplifies bremsstrahlung losses, so that purely thermonuclear ignition would require ion temperatures of several hundred keV. Nonthermal schemes employing high-intensity laser-driven proton beams~\cite{PhysRevE.72.026406, Labaune:2013aa, PhysRevX.4.031030, PhysRevE.103.053202, Margarone:2022aa} bypass this barrier, and they share the same physics with proton fast ignition~\cite{Roth2001FastIgnition, Fernandez2014FastIgnition}, where a short proton pulse must deposit a known fraction of its energy inside compressed deuterium--tritium fuel. In both cases the reaction yield and the deposition profile depend explicitly on the proton energy distribution \emph{inside} the target.

Yields of laser-driven p--B reactions have most often been inferred from CR-39 or imaging-plate detection of the emitted $\alpha$ particles~\cite{Cartwright1978, Cassou1978, Sonoda1983, Amemiya1988, Chen2011}. Track-detector approaches are well calibrated for monoenergetic accelerator beams~\cite{RUSETSKII2025165651}, but they encounter several systematic difficulties at high-intensity laser facilities. Surface contaminants such as carbon and oxygen are readily co-accelerated; heavy ions other than $\alpha$'s leave tracks of overlapping diameter~\cite{10.1063/1.4927684}, blurring the species identification; and the relativistic plasma generates strong sheath electric and self-generated magnetic fields~\cite{PhysRevLett.70.3059} that distort both the angular distribution and the escaping fraction of $\alpha$'s. Cross-checking with external proton diagnostics is no easier: Thomson parabolas and radiochromic film stacks~\cite{Carroll2010, Kantarelou2023, Nurnberg2009, Abe2021b} record only ions that escape the rear surface, and only after the same sheath fields have shaped their trajectories. Neither approach gives direct access to the in-solid proton energy distribution responsible for the bulk of the nuclear reactions, which motivates the activation strategy adopted in this work.

% -----------------------------------------------------------------------
% Appendix B (former Sec. SM-2)
% -----------------------------------------------------------------------
\section{Appendix B: Bayesian evaluation of nuclear reaction cross sections}
\label{app:Bayesian}
 
The nuclear cross sections relevant to this work---$\mathrm{^{10}B}(p,\alpha)\mathrm{^{7}Be}$, $\mathrm{^{11}B}(p,n)\mathrm{^{11}C}$, and $\mathrm{^{11}B(p,2\alpha)^{4}He}$---are imperfectly known across the energy range of interest. 
Reported $\mathrm{^{11}B}(p,n)\mathrm{^{11}C}$ values between 4 and 10~MeV differ between experiments by factors of several, producing a comparably large statistical dispersion. 
To incorporate these uncertainties consistently into the activation analysis, we evaluate each cross section as the posterior of a Bayesian model fit to the experimental datasets summarized in Table~\ref{tab:xs_list}.
 
We model the logarithmic cross section $\log\sigma(E)$ as a smooth latent function of $\log E$ and use a Student-$t$ observation model (specified below) to absorb dataset-to-dataset disagreement.
Posterior reconstructions are exported on the energy range relevant to the activation analysis, $E_{\mathrm{cm}}\le 30$~MeV: above this energy the laser-accelerated proton energy distribution contributes negligibly to the activation yield, and experimental cross-section data become too sparse to constrain the posterior meaningfully.
 
For each observation $i$, the inputs are the center-of-mass energy $E_{\mathrm{cm},i}$ and the observed cross section $\sigma^{\mathrm{obs}}_{i}$.
The absolute uncertainty $\delta_i$ is taken as the reported total uncertainty when available; when statistical and systematic uncertainties are reported separately, we use their quadrature sum.
For data points with no reported uncertainty, we assign a default relative error and additionally impose a lower bound on the relative error to prevent vanishingly small uncertainties from dominating the likelihood.
The analysis is performed in logarithmic space using
\begin{equation}
x_i = \log E_{\mathrm{cm},i}, \qquad
y_i = \log \sigma^{\mathrm{obs}}_{i}, \qquad
s_i = \log\!\left(1+\frac{\delta_i}{\sigma^{\mathrm{obs}}_{i}}\right)
\end{equation}
as the observation variables.
 
The latent log cross section $f(x)=\log\sigma(x)$ is represented as the sum of a P-spline on log-energy and a high-energy tail term,
\begin{equation}
f(x)=\sum_{k=1}^{K}\beta_k B_k(x) + a\,h(x) + b\,h(x)^2,
\qquad
h(x)=\max(0, x-x_{\mathrm{tail}})
\end{equation}
where $B_k(x)$ are B-spline basis functions; we use $K=24$ cubic basis functions for both reactions.
The tail break point $x_{\mathrm{tail}}$ is determined from the highest center-of-mass energy in the existing data.
The smoothness of the spline coefficients is constrained by a prior on the second-order difference,
\begin{equation}
\Delta^2\beta_k = \beta_{k+2}-2\beta_{k+1}+\beta_k,
\end{equation}
of the form
\begin{equation}
\Delta^2\beta_k \sim \mathcal{N}(0,\tau_{\mathrm{smooth}}^2),
\qquad
\tau_{\mathrm{smooth}}\sim \mathrm{HalfNormal}(0.5).
\end{equation}
 
A Student-$t$ likelihood is adopted for the observation model,
\begin{equation}
y_i \sim \mathrm{StudentT}_{\nu}\!\left(\mu_i, s_i^{\mathrm{tot}}\right),
\end{equation}
with the total scale parameter defined as
\begin{equation}
\left(s_i^{\mathrm{tot}}\right)^2
=
s_i^2 + \tau_{\mathrm{int}}^2 + \lambda(x_i)^2 + \left(\tau_{\mathrm{tail}} d_i\right)^2.
\end{equation}
Here $\tau_{\mathrm{int}}$ is an additional dispersion shared by all observations, $\lambda(x)$ is a locally varying dispersion interpolated on log-energy, and $\tau_{\mathrm{tail}}\,d_i$ represents an additional uncertainty in the tail region with $d(x)=\max(0,x-x_{\mathrm{tail}})/\log(E_{\max}/E_{\mathrm{tail}})$ a normalized tail distance and $d_i=d(x_i)$.
We assign $\tau_{\mathrm{int}}\sim\mathrm{HalfNormal}(0.25)$ and $\tau_{\mathrm{tail}}\sim\mathrm{HalfNormal}(\sigma_{\mathrm{tail}})$.
The local dispersion is constructed by linearly interpolating $\lambda_j\sim\mathrm{HalfNormal}(\sigma_\lambda)$ defined at six knots on log-energy; the prior scales $\sigma_\lambda$ and $\sigma_{\mathrm{tail}}$ are specified separately for each reaction below.
The degrees of freedom are fixed at $\nu=20$ for both reactions.

\begin{table}[htbp]
    \centering
    \caption{Experimental datasets used for the Bayesian inference of the reaction cross sections.}
    \label{tab:xs_list}
    
    \begin{tblr}{
    colspec = {X c X},
    rows    = {font=\footnotesize},
    rowsep  = 0pt,
    stretch = 0.9,
    hline{1,Z} = {0.8pt},
    hline{2}   = {0.5pt},
    row{1}     = {font=\small\bfseries},
    }
    
    Reference & Year & Source \\
    
    \hline
    \SetCell[c=3]{l}{\bfseries $^{11}$B(p,n)$^{11}$C} \\
    
    Anders \textit{et al.}~\cite{Anders:1981aa} & 1981 & EXFOR A0330002 \\
    Ramavataram \textit{et al.}~\cite{Ramavataram:1980aa} & 1980 & EXFOR T0041002 \\
    Sklavenitis~\cite{Sklavenitis1966} & 1966 & EXFOR O2133002 \\
    Segel \textit{et al.}$^{a}$~\cite{PhysRev.139.B818} & 1965 & EXFOR F0332006 \\
    Valentin \textit{et al.}~\cite{VALENTIN196581} & 1965 & EXFOR C0062004 \\
    Legge \textit{et al.}~\cite{LEGGE1961616} & 1961 & EXFOR F0283002 \\
    Furukawa \textit{et al.}~\cite{doi:10.1143/JPSJ.15.2167} & 1960 & EXFOR P0045002 \\
    Gibbons \textit{et al.}$^{a}$~\cite{PhysRev.114.571} & 1959 & EXFOR T0010003 \\
    Kalinin \textit{et al.}~\cite{Kalinin1957_SJA} & 1957 & EXFOR A0923004 \\
    Blaser \textit{et al.}$^{a}$~\cite{Blaser1951} & 1951 & EXFOR D0095002 \\
    
    \hline
    \SetCell[c=3]{l}{\bfseries $^{10}$B(p,$\alpha$)$^{7}$Be} \\
    
    Szab\'o \textit{et al.}~\cite{SZABO1972527} & 1972 & EXFOR F0345002 \\
    Valentin \textit{et al.}~\cite{VALENTIN1963163} & 1963 & EXFOR C0061005 \\
    Kalinin \textit{et al.}~\cite{Kalinin1957_SJA} & 1957 & EXFOR A0923003 \\
    Bach \textit{et al.}~\cite{Bach01081955} & 1955 & EXFOR O0917002, O0917003 \\
    Caciolli \textit{et al.}$^{b}$~\cite{caciolli_2016} & 2016 & Table~1 \\
    Kafkarkou \textit{et al.}$^{c}$~\cite{KAFKARKOU201348} & 2013 & Table~1 \\
    Tian \textit{et al.}$^{d}$~\cite{PhysRevC.110.045806} & 2024 & Table~II \\
    Vande Kolk \textit{et al.}$^{e}$~\cite{PhysRevC.105.055802} & 2022 & digitized data \\
    
    \end{tblr}
    
    \vspace{2pt}
    {\footnotesize
    
    $^{a}$ Data points with $E_{\mathrm{cm}} < 2.77~\mathrm{MeV}$ reported in the datasets of Segel, Gibbons, and Blaser were excluded because they lie below the physical threshold of the $^{11}$B(p,n)$^{11}$C reaction ($Q \approx -2.77~\mathrm{MeV}$).
    
    $^{b}$ Cross sections reconstructed from the tabulated astrophysical S-factor values in Table~1.
    
    $^{c}$ Total cross sections derived in the original work from angular differential cross sections using a Legendre polynomial fit ($\sigma_{\mathrm{tot}} = 4\pi A_0$).
    
    $^{d}$ Cross sections reconstructed from the S-factor values reported in Table~II.
    
    $^{e}$ Differential cross sections were converted to total cross sections using the Legendre polynomial expansion and summed over reaction channels $\alpha_0$ and $\alpha_1$.
    }
\end{table}
 
Posterior sampling was performed in PyMC~\cite{pymc2023} using the No-U-Turn Sampler with multiple chains. Convergence was diagnosed with the Gelman--Rubin statistic ($\hat{R}\lesssim 1.01$ for the main parameters) and a sufficient effective sample size, both confirming reliable mixing. 
The target acceptance was set to 0.98 and the random seed to 20260413.
The continuous cross section is evaluated on an 800-point grid spanning $0.01$--$30$~MeV; on this grid, the cross section in the region below the lowest fitted data point is set to zero.
 
For each posterior sample $s$, we compute
\begin{equation}
    f^{(s)}(x), \qquad \sigma^{(s)}(x)=\exp\{f^{(s)}(x)\}
\end{equation}
on the grid and construct the latent curves from their quantiles.
In addition, to accommodate inconsistencies among datasets in the credible bands, we adopt an envelope-mode construction.
The base envelope is defined as
\begin{equation}
    e_{\mathrm{base}}(x)= \sqrt{\mathbb{E}\!\left[(c\,\lambda(x))^2\right] + \mathbb{E}\!\left[(\tau_{\mathrm{tail}}d(x))^2\right]},
\end{equation}
where $c$ is a coefficient that amplifies the contribution of the local dispersion.
In addition, using the posterior median $\tilde{f}(x)$ of the latent curve, we construct a residual envelope from the observation residuals
\begin{equation}
    r_i = \left|y_i-\tilde{f}(x_i)\right|
\end{equation}
through weighted quantiles,
\begin{equation}
e_q(x)=Q_w\!\left(q;\{r_i\},w_i(x)\right),
\qquad
w_i(x)=\exp\!\left[-\frac{(x_i-x)^2}{2\ell^2}\right].
\end{equation}
For both reactions we use $\ell=0.08$, $q_{68}=0.68$, and $q_{95}=0.95$, with an additional factor of 1.08 applied to the 95\% band.
The envelope at each credible level is taken as the pointwise maximum of $e_{\mathrm{base}}(x)$ and $e_q(x)$, and the final 68\% and 95\% bands are obtained by widening the corresponding latent credible bands by this envelope on each side.
 
\textit{$^{10}\mathrm{B}(p,\alpha)\mathrm{^{7}Be}$.}~
For data points without reported uncertainties, we assign a default relative error of 20\%, with a lower bound of 3\%.
To absorb dataset-to-dataset normalization differences, we redefine the observation mean as
\begin{equation}
    \mu_i = f(x_i) + \eta_{g(i)},
\end{equation}
introducing a per-dataset normalization offset $\eta_g$ as an auxiliary parameter.
The offsets are centered to zero mean across datasets, $\sum_g \eta_g = 0$, to ensure identifiability with the absolute scale of $f(x)$.
The prior width of $\eta_g$ is set from the median relative systematic uncertainty within each dataset, with 15\% as the default fallback and a 5\% floor.
We set $(\sigma_\lambda, \sigma_{\mathrm{tail}}) = (1.2, 1.0)$.
For the conservative credible-band construction, the local-dispersion amplification factor is $c=2.0$ and the final smoothing width of the residual envelope is 0.20 on log-energy.
The minimum and maximum energies of the fit are not fixed; the data points are used as they are.
 
\textit{$^{11}\mathrm{B}(p,n)\mathrm{^{11}C}$.}~
Because this is a threshold reaction, data points with $E_{\mathrm{cm}}< 2.77$~MeV are excluded from the fit, and the upper energy limit of the fit is set to 40~MeV.
For data points without reported uncertainties, we assign a default relative error of 25\%, with a lower bound of 5\%.
For this reaction, no inter-dataset normalization offset is introduced, and the observation mean is simply
\begin{equation}
    \mu_i = f(x_i).
\end{equation}
We set $(\sigma_\lambda, \sigma_{\mathrm{tail}}) = (0.45, 0.25)$.
For the conservative credible-band construction, the local-dispersion amplification factor is $c=1.35$ and the final smoothing width of the residual envelope is 0.08 on log-energy.

% -----------------------------------------------------------------------
% Appendix C (former Sec. SM-3)
% -----------------------------------------------------------------------
\section{Appendix C: Activation measurement, nuclide identification, and yield evaluation}
\label{sec:SM_exp_details}

This section details the activation measurement procedure, including target and laser conditions, nuclide identification, spatial-distribution analysis, detection-efficiency correction, and the conversion from observed counts to absolute production yields. Representative examples of $\gamma$-ray spectra and decay-curve fits are shown in Figs.~\ref{fig:spectrum} and~\ref{fig:Activation_curve}.

The per-shot laser parameters and target materials are summarized in Table~\ref{tab:experiment_list}.

\begin{table*}[ht]
\centering
\caption{Per-shot laser parameters and target materials. The on-target intensity quoted in the main text is evaluated from the on-target energy, pulse width, and the spot size listed below.}
\label{tab:experiment_list}
\begin{tblr}{
    colspec = {l l c c},
    rows    = {font=\small},
    rowsep  = 1pt,
    stretch = 1.05,
    hline{1,Z} = {0.8pt},
    hline{2}   = {0.5pt},
    row{1}     = {font=\small\bfseries},
}
Shot ID & Target material & On-target energy (J) & Pulse width (ps) \\
\SetCell[c=4]{l}{\bfseries in-target scheme} & & & \\
L5944 & Decaborane & $800 \pm 20$ & $1.5$ \\
L5947 & Borophane  & $540 \pm 20$ & $0.9$ \\
L5949 & Decaborane & $540 \pm 20$ & $1.0$ \\
\SetCell[c=4]{l}{\bfseries pitcher--catcher scheme} & & & \\
L5941 & Borophane  & $710 \pm 20$ & $1.6$ \\
L5945 & Decaborane & $780 \pm 20$ & $1.8$ \\
L5948 & Borophane  & $560 \pm 20$ & $0.9$ \\
\end{tblr}
\end{table*}

\subsection{Aluminum-foil collector and self-activation}
The inner surface of the cylindrical debris collector (purity~$\ge$99\%, foil thickness 100~$\mu$m) was lined with removable aluminum foil to trap radioactive debris and to suppress unintended activation of the collector body itself. After each shot the foil was extracted and mounted on a holder of identical material and geometry that had not been laser-irradiated, ensuring that the measured radioactivity originated exclusively from the deposited debris.

The dominant aluminum activation channel that could in principle contaminate the 511~keV measurement is $\mathrm{^{27}Al}(p,x)\mathrm{^{18}F}$ ($T_{1/2}\approx 100$~min, peak cross section $\sim 10$~mb, effective threshold $35\text{--}40\,\mathrm{MeV}$)~\cite{LAGUNASSOLAR198841}.
Under the present conditions, very few protons reach this energy, and the foil subtended only a small solid angle relative to the proton emission cone. Foil self-activation is therefore negligible.

\subsection{Time-separated $\gamma$-ray spectroscopy}
Two coaxial high-purity germanium (HPGe) detectors (Canberra GC4018 and GC3226, relative efficiencies 40\% and 32\% with respect to a 3-inch $\times$ 3-inch NaI standard) were used. Their full-width-at-half-maximum energy resolutions at the 1.33~MeV $^{60}\mathrm{Co}$ line were 1.8 and 2.6~keV, respectively. Each detector was housed in a 10~cm-thick lead shield to suppress background; the 76.2~mm aluminum endcap of each detector faced the laser-incidence side of the debris collector.

\begin{figure}[htbp]
    \centering
    \includegraphics[width=1.0\linewidth]{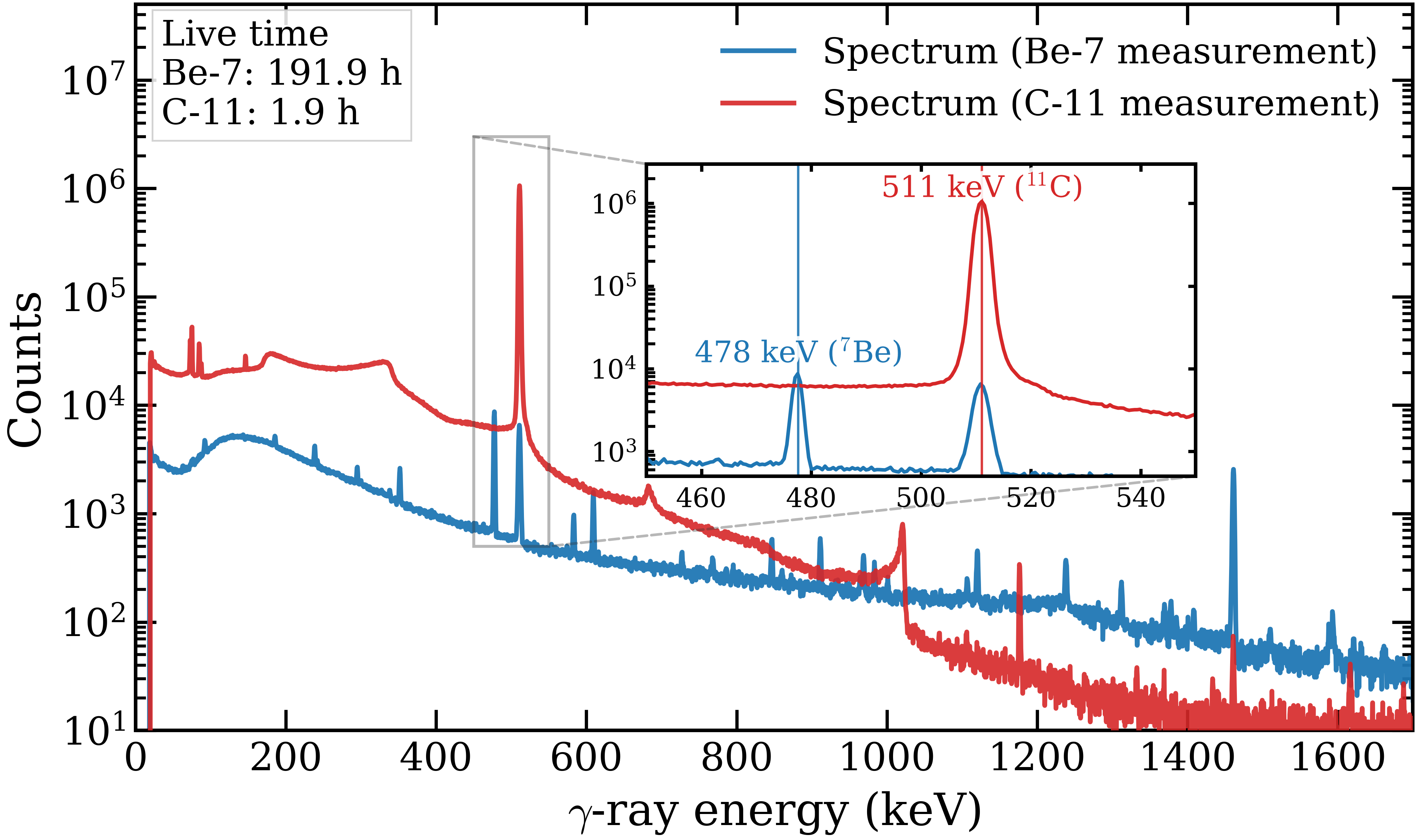}
    \caption{
Representative $\gamma$-ray spectrum obtained in the pitcher--catcher configuration (shot~L5948). The 511~keV and 478~keV peaks correspond to the decay of $^{11}\mathrm{C}$ and $^{7}\mathrm{Be}$, respectively.
    }
    \label{fig:spectrum}
\end{figure}

The 20.4~min half-life of $^{11}\mathrm{C}$ and 53.2~day half-life of $^{7}\mathrm{Be}$~\cite{ENSDF} preclude simultaneous quantification within a single counting window. The $^{11}\mathrm{C}$ measurement was therefore conducted from $\sim 20$~min to $\sim 1$~h after each shot, and the $^{7}\mathrm{Be}$ measurement from $\sim 3$~h to $\sim 7$~days. In all six shots, pronounced peaks at $511.1\pm 1.5$~keV and $477.7\pm 0.8$~keV were observed (Fig.~\ref{fig:spectrum}), corresponding to the annihilation $\gamma$ following $\beta^{+}$ decay of $^{11}\mathrm{C}$ and to the de-excitation $\gamma$ following electron capture of $^{7}\mathrm{Be}$ to the first excited state of $^{7}\mathrm{Li}$.

\subsection{Decay-curve analysis and impurity subtraction}
In short-time activation measurements, the peak counting rate $B(t)$ at 511~keV cannot be reproduced by a single $^{11}\mathrm{C}$ component, and a multi-component fit accounting for three $\beta^{+}$-emitting nuclides is required. The nuclides considered are as follows.
\begin{enumerate}
\item[(i)] $^{11}\mathrm{C}$ ($T_{1/2}=20.4$~min): the product nuclide of the target reaction $^{11}\mathrm{B}(p,n)^{11}\mathrm{C}$ in this work.
\item[(ii)] $^{34\mathrm{m}}\mathrm{Cl}$ ($T_{1/2}=32.0$~min, $\lambda=3.61\times10^{-4}$~s$^{-1}$~\cite{ENSDF}): produced by proton reactions on chlorine- and/or sulfur-containing contaminants introduced during target fabrication and handling~\cite{Vandecasteele:1980aa, Lagunas-Solar:1992aa}. Because this nuclide also emits several characteristic $\gamma$-ray lines other than the 511~keV annihilation peak (146.4, 1176.6, 2127.5, and 3304.0~keV), its abundance can be estimated independently from the intensities of these other peaks. We use the abundance estimated from these other peaks as the initial guess in the fit.
\item[(iii)] $^{13}\mathrm{N}$ ($T_{1/2}=9.97$~min): produced by proton reactions on surface C/N/O contamination, $^{12}\mathrm{C}(p,\gamma)^{13}\mathrm{N}$, $^{13}\mathrm{C}(p,n)^{13}\mathrm{N}$, $^{14}\mathrm{N}(p,d)^{13}\mathrm{N}$, and $^{16}\mathrm{O}(p,\alpha)^{13}\mathrm{N}$. Because of its short half-life of about 10 minutes, its contribution to the 511~keV peak is non-negligible, particularly during the early stages of short-time measurements.
\end{enumerate}

\begin{figure}[htbp]
    \centering
    \subfloat[\label{fig:C11_Identification}]{%
        \includegraphics[width=0.48\linewidth]{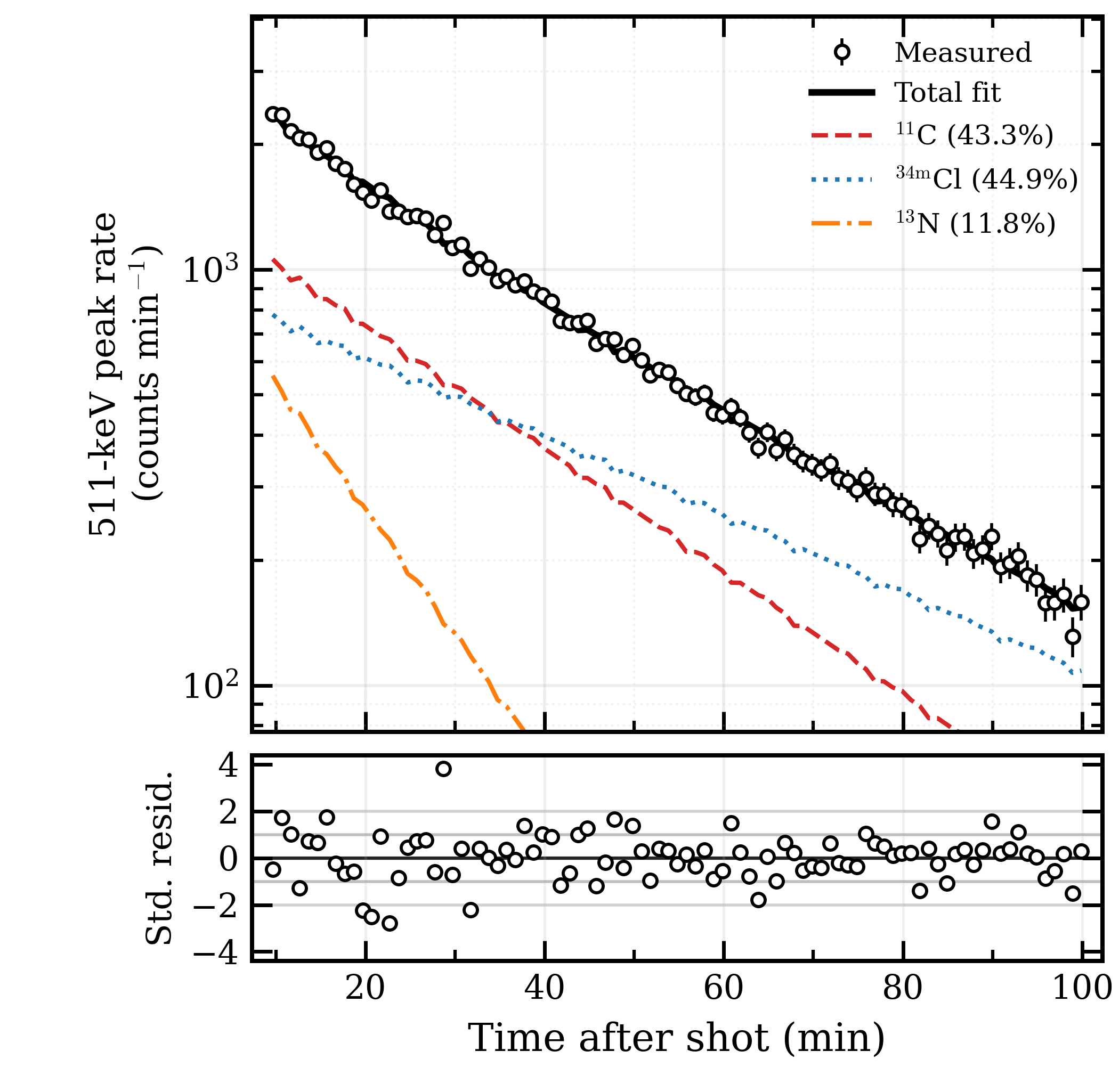}
    }
    \hfill
    \subfloat[\label{fig:Be7_Identification}]{%
        \includegraphics[width=0.48\linewidth]{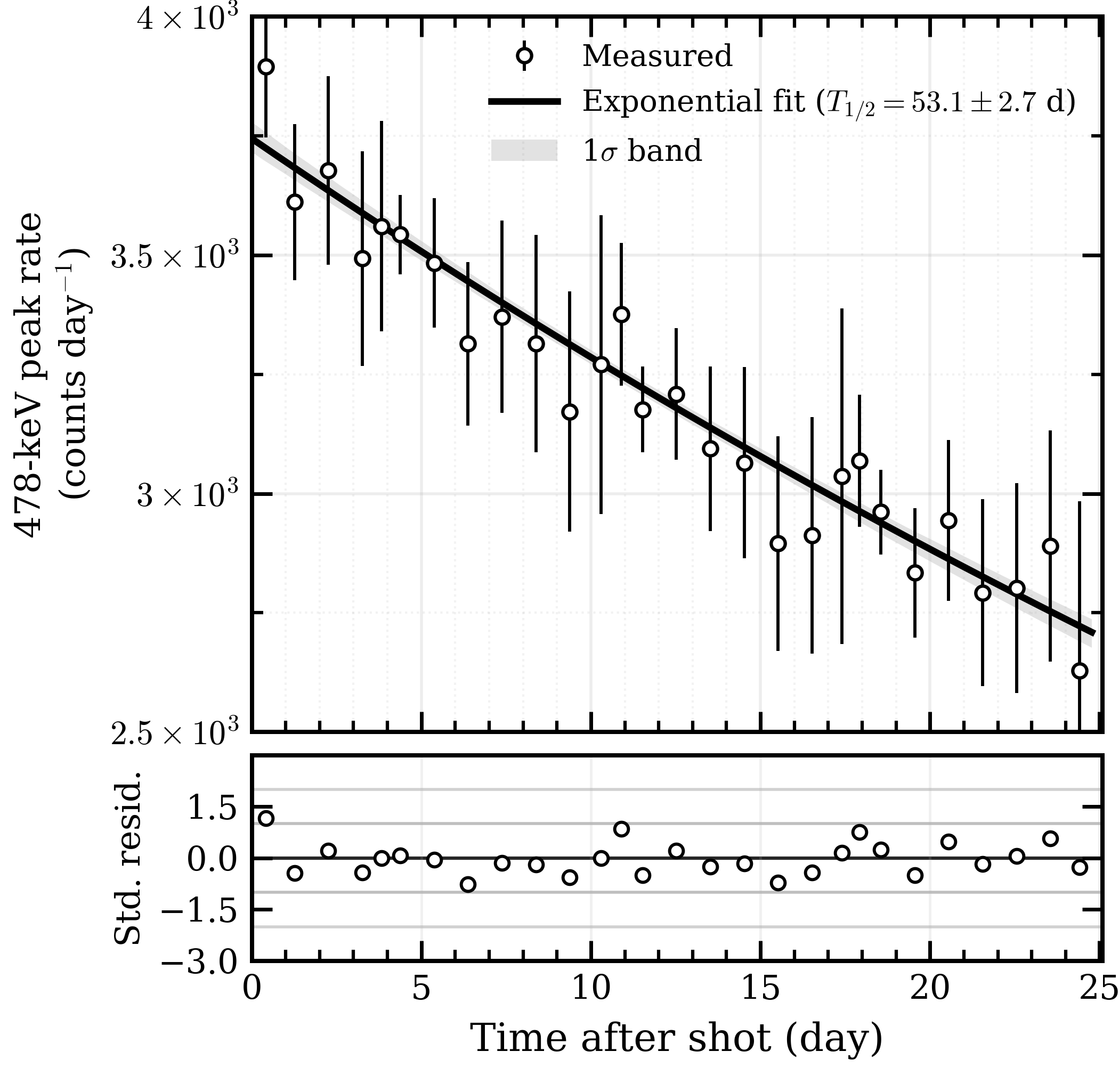}
     }
    \caption{
Decay-curve analysis used for nuclide identification.
(a)~Temporal evolution of the 511~keV peak count rate for shot~L5949.
Open circles show the measured count rate with counting-statistics error bars.
The black curve denotes the total fit, while the red dashed, blue dotted, and orange dash-dotted curves represent the fitted $^{11}\mathrm{C}$, $^{34\mathrm{m}}\mathrm{Cl}$, and $^{13}\mathrm{N}$ components, respectively.
The percentages in the legend indicate the fractional contributions of these radioactive components over the fitted time window.
A constant offset term is included in the fit but omitted from the plot for clarity.
(b)~Temporal evolution of the 478~keV peak count rate and a single-exponential fit, yielding $T_{1/2}=53.1\pm2.7$~day, consistent with the literature value for $^{7}\mathrm{Be}$ (53.2~day~\cite{ENSDF}).
The shaded band denotes the 1$\sigma$ fit uncertainty. The lower panels in (a) and (b) show the standardized residuals of the fits.
    }
    \label{fig:Activation_curve}
\end{figure}

Accordingly, the counting rate is fitted by the sum of three exponential components and a constant offset term,
\begin{equation}
\begin{aligned}
B(t) ={}& B_{\mathrm{bg}} \\
&+ B_{(\ce{^{11}C})}\exp\!\left[-\lambda_{(\ce{^{11}C})}t\right] \\
&+ B_{(\ce{^{34m}Cl})}\exp\!\left[-\lambda_{(\ce{^{34m}Cl})}t\right] \\
&+ B_{(\ce{^{13}N})}\exp\!\left[-\lambda_{(\ce{^{13}N})}t\right],
\end{aligned}
\end{equation}
as illustrated in Fig.~\ref{fig:Activation_curve}(a).
The decay constants $\lambda$ of each nuclide are fixed at their literature values~\cite{ENSDF}, whereas the three amplitudes $B_{(\cdot)}$ and the constant offset term $B_{\mathrm{bg}}$ are treated as free parameters.
For $^{34\mathrm{m}}\mathrm{Cl}$, providing the independent estimate from other $\gamma$-ray peaks as the initial guess (as described above) improves the stability and convergence of the fit.

The $^{11}\mathrm{C}$ fraction $f_{^{11}\mathrm{C}}$ obtained from the fit differs markedly between the two configurations: 0.9--1.0 for the pitcher--catcher case versus 0.3--0.4 for the in-target case.
The lower $^{11}\mathrm{C}$ fraction in the in-target case is consistent with the lower proton energies observed in the Thomson-parabola measurement on the rear side of the target.
In this configuration, fewer protons exceed the reaction threshold for $^{11}\mathrm{B}(p,n)^{11}\mathrm{C}$ ($E_{\mathrm{cm}}=2.77~\mathrm{MeV}$), reducing the absolute production of $^{11}\mathrm{C}$.
The reduced absolute production of $^{11}\mathrm{C}$ increases the relative contribution of contamination-related components such as $^{34\mathrm{m}}\mathrm{Cl}$ and $^{13}\mathrm{N}$ to the 511~keV peak.
The values of $f_{^{11}\mathrm{C}}$ shown in Table~\ref{tab:summary_results} are those obtained from this three-component fit.

The 478~keV peak, by contrast, is well described by a single exponential with fitted half-life $53.1\pm 2.7$~day, consistent with the $^{7}\mathrm{Be}$ literature value (53.2~day~\cite{ENSDF}; Fig.~\ref{fig:Activation_curve}(b)); no other contributing nuclides were identified in this energy region, and we set $f_{^{7}\mathrm{Be}}=1.0$.

\subsection{Spatial distribution, efficiency, and absolute yields}
At early times the 511~keV signal is dominated by $^{11}\mathrm{C}$, so the autoradiographic image primarily reflects the spatial distribution of $^{11}\mathrm{C}$ on the collector inner surface; we assume the same spatial pattern for $^{7}\mathrm{Be}$, which is co-produced and co-transported as debris. The position-dependent HPGe detection efficiency was computed by Geant4 Monte Carlo simulations~\cite{Agostinelli:2003aa} and absolutely calibrated against a $^{152}\mathrm{Eu}$ and $^{133}\mathrm{Ba}$ standard source~\cite{Sarasti:2022}.

The number of nuclei of isotope ``iso'' that decayed during the measurement window $[t_{\mathrm{s}},t_{\mathrm{e}}]$ is then
\begin{equation}
N_{\mathrm{dec}}^{(\mathrm{iso})} = \frac{f_{\mathrm{iso}}\, N_{\mathrm{cnt}}}{I_{\gamma}^{(\mathrm{iso})}\,\sum_{i}\varepsilon_{i} P_{i}},
\end{equation}
where $N_{\mathrm{cnt}}$ is the total peak count, $I_{\gamma}^{(\mathrm{iso})}$ is the $\gamma$ emission probability per decay (with the factor $1/2$ applied for $^{11}\mathrm{C}$ since one of the two annihilation $\gamma$'s escapes detection on average), and $\varepsilon_{i}$ and $P_{i}$ are the position-resolved detection efficiency and normalized debris distribution at pixel $i$. The corresponding production yield at the moment of laser irradiation is
\begin{equation}
N_{0}^{(\mathrm{iso})} = \frac{1}{\eta_{\mathrm{rec}}}\,\frac{N_{\mathrm{dec}}^{(\mathrm{iso})}}{\exp(-\lambda_{\mathrm{iso}} t_{\mathrm{s}})-\exp(-\lambda_{\mathrm{iso}} t_{\mathrm{e}})},
\end{equation}
with $\eta_{\mathrm{rec}}=0.96$ the geometric debris-collection efficiency and $\lambda_{\mathrm{iso}}$ the decay constant from ENSDF~\cite{ENSDF}. The resulting $N_{0}^{(^{11}\mathrm{C})}$ and $N_{0}^{(^{7}\mathrm{Be})}$ values for each shot are listed in Table~\ref{tab:summary_results}.

% =======================================================================
% Bibliography
% Single bibliography for the entire document (main text + appendices).
% Duplicate \cite keys are automatically deduplicated by BibTeX.
% =======================================================================
\bibliography{reference.bib}

\end{document}